\documentclass[twocolumn]{revtex4}
\usepackage{amssymb}
\usepackage{tabularx}
\usepackage{xcolor}
\usepackage{amsmath}

\def\ga{\gamma}

\def\eps{\varepsilon}

\def\la{\lambda}
\def\si{\sigma}

\def\La{\Lambda}
\def\la{\lambda}

\def\cL{{\mathcal L}}

\def\cN{{\mathcal N}}
\def\cO{{\mathcal O}}

\def\11{{\mathbb 1}}

\def\re{{\rm e}}

\def\rd{{\rm d}}

\def\keV{\,\rm keV}

\def\TeV{\,\rm TeV}

\def\beq{\begin{equation}}
\def\eeq{\end{equation}}
\def\bea{\begin{eqnarray}}
\def\eea{\end{eqnarray}}
\def\nn{\nonumber}

\begin{document}
\title{Supermassive charged gravitinos and the Lux-Zeplin event}
\author{Krzysztof A. Meissner$^1$ and Hermann Nicolai$^2$\\}
\affiliation{\\
$^1$Faculty of Physics,
University of Warsaw\\
Pasteura 5, 02-093 Warsaw, Poland\\
$^2$Max-Planck-Institut f\"ur Gravitationsphysik
(Albert-Einstein-Institut)\\
M\"uhlenberg 1, D-14476 Potsdam, Germany
}

\vspace{10mm}

\begin{abstract}  The Lux-Zeplin collaboration recently reported one unusual event 
which was interpreted as first evidence for a WIMP (arXiv:2609.02823[hep-ex]\cite{LZ1}). 
In this note we offer an alternative explanation of this event, following our earlier 
proposal that dark matter could consist of fractionally charged supermassive 
gravitinos. A main difference between the two interpretations is that one expects
small cross sections and large flux rates for WIMPs, whereas it is the opposite for 
supermassive gravitinos, namely large (electromagnetic) cross sections and 
very low flux rates. The observed nuclear recoil energy of about 
250 keV and the occurrence of one event in one year fit well 
with the estimates quoted in (K.A. Meissner, H.~Nicolai,
Eur.Phys.J.C 84 (2024) 3, 269,\cite{MNLZ}). A main characteristic of the
new interpretation distinguishing it unequivocally from the WIMP hypothesis 
would be the sequential occurrence of two or more recoils 
within a few microseconds along the gravitino's straight trajectory, 
a prediction that remains to be checked. Such a measurement
would also allow to overcome the `neutrino floor'. In this note we refine our 
previous estimates and propose some extra steps so as to enable the experiment 
to discriminate between the WIMP and gravitino hypotheses.
\end{abstract}
\maketitle

\vspace{5mm}

\section{Introduction.} 
The nature of dark matter (DM) remains one of the enduring
mysteries of modern astroparticle physics, see \cite{Strumia} for
an extensive overview of the theoretical and experimental aspects
and a review of the currently favored models. Somewhat outside
the generally accepted possible scenarios (axions and ALPs, WIMPs, 
low energy supersymmetry, {\em etc.}) we have argued in \cite{MeissnerNicolai2019} 
for the unconventional possibility that DM could consist at least in part of 
an extremely dilute gas of supermassive stable gravitinos with charge 
$q=\pm\frac23 e$. This proposal is chiefly motivated by the 
remarkable agreement between a unification scenario linked to, 
but not directly based on, $\cN=8$ supergravity and the 
observed fermion spectrum of the Standard Model of particle
physics with 48 = 3$\,\times\,16$ quarks and leptons \cite{GM,MeissnerNicolai2015,JHEP}. 
An unusual feature of this 
proposal is that, due to their large mass and very low abundance such 
DM candidates can carry $\cO(1)$ electric charges and would thus not 
be dark at all, but luminous.  Such particles can escape detection 
only by virtue of their extremely rare occurrence. 

The Lux-Zeplin (LZ) collaboration \cite{Xenon,LZ} very recently reported 
one outlier event \cite{LZ1} that was interpreted as first possible
direct evidence for a WIMP (= weakly interacting massive particle), 
one of the main candidates for explaining
dark matter. More precisely, the measured event consists of a  
nuclear recoil of $ 248 \pm 23 (stat) \pm 23 (sys) \keV$, and was observed
in a region of parameter space where the known background expectation 
is low,  during an observation period of 220 days of data collected 
between 27 March 2023 and 1 April 2024 \cite{LZ1}. 
To be sure, the data by themselves 
do not allow to identify what kind of particle was observed. 
Lacking a definite DM model (especially in view of the no-show of low 
energy supersymmetry) \cite{LZ1} resorts to the (non-relativistic) 
effective field theory (EFT) methods developed in \cite{EFT1,EFT2}  
which are largely `agnostic' {\em w.r.t.} the precise nature of DM.

For an interpretation in terms of WIMPs and effective field theory, the 
explanation is modeled in terms of a four Fermi interaction, with 
nucleons $N$ and new hypothetical fermionic DM candidates $\chi$,
of the schematic form 
\beq\label{chiN}
\cL_{\rm WIMP} \,\propto \, \La^{-2}\, \bar\chi\chi\, \bar N N
\eeq
with an unknown coupling $\La$ of dimension mass (there 
are various possible kinematical structures in (\ref{chiN}) \cite{EFT1,EFT2}).
With an assumed WIMP mass of $\lesssim \cO(1 \TeV)$
the flux rate is about $10^{15}\,$m$^{-2}$yr$^{-1}$sr$^{-1}$ 
(assuming that all of DM is made up of
such WIMPs), and the corresponding WIMP-nucleon cross section must therefore be 
very small, on the order of $\lesssim 10^{-46}\,$m$^2$, allowing for at most
one interaction of the DM particle in the Xenon detector (the fiducial volume
of the LZ detector is about 2\,m$^3$). Like other DM scenarios,
this scenario thus relies on several critical assumptions, in particular the 
existence of new types of (super-)weak interactions, for instance 
mediated by a new photon $A'_\mu$, of mass $\sim\La$, 
that must mix very weakly with standard electromagnetism.
In that case the model parameters, and thus the cross section, 
must be adjusted in accordance with measured data. As an
inspection of the ${\tt arXiv}$ shows, there is by now already a considerable 
number of proposals based on different WIMP-type
ans\"atze to explain the LZ event~\footnote{There are also
some attempts to explain the event in terms of neutrino or other backgrounds.}, whose common feature is an 
assumed WIMP mass on the order of the TeV scale 
and very small cross section.

In this note we wish to present an alternative explanation that
goes in a very different direction, and is based on 
the DM proposal made in \cite{MeissnerNicolai2019}, which relies 
on supermassive electrically charged gravitinos (in the remainder 
we will usually refer to the hypothetical supermassive gravitino 
simply as `the Particle'). In this case the gravitino-nucleon 
four fermion interaction is uniquely fixed and 
follows simply from the usual coupling of the gravitino to 
the electromagnetic potential $A_\mu$, {\em viz.}
\beq\label{Lint}
\cL_{int} \, = \, q\, \bar \psi_\mu \ga^{\mu\nu\rho} A_\nu \psi_\rho
\eeq 
where $\psi_\mu$ is the (massive) gravitino vector spinor and $q$ the 
gravitino charge. Therefore, neglecting spin effects 
in the non-relativistic approximation, the cross section between
the putative DM particle and the Xenon atoms in the detector is large, 
namely electromagnetic, and well described by the Lindhard formula \cite{Lindhard}.
Assuming a gravitino mass close to the Planck mass
the low event rate is then solely due to an extremely small flux, 
which is smaller by a factor $\sim \cO(10^{-15})$ than the assumed 
flux for WIMPs. Some relevant estimates were already presented in \cite{MNLZ}.
We emphasize that in this case and in contrast to the WIMP hypothesis, the 
flux is the {\em only} free parameter, while the cross section is fixed at the standard
value $\sim 10^{-28}\,$m$^2$. Remarkably, under these restricted 
conditions, the event rate and the nuclear recoil energy quoted in \cite{MNLZ} 
for the LZ detector (indeed before the release of the recent LZ results!)
are in reasonable agreement with the findings of \cite{LZ1}, keeping 
in mind various uncertainties such as the DM density and velocity distribution 
in the vicinity of the solar system.

Consequently, the distinguishing feature of our proposal is that gravitinos 
react electromagnetically both with the Xenon nuclei and the electron clouds
once they enter the detector (which they can do from any
direction), leaving a clear trace of their passage through the detector.
In particular, even two or more sequential nuclear recoils within a few 
microseconds could be observed, in marked contrast to WIMP induced 
events. To find out whether multiple recoils actually occur the challenge 
would be to discriminate a one recoil event from a two recoil event 
within a time span of only a few microseconds, which is the 
time which the particle would be expected to spend in the detector.

\section{Interactions of an electrically charged supermassive particle}

We briefly recall the basic features of our proposal, which has its origins in 
Gell-Mann's observation \cite{GM} that, subject to a `spurion shift' of the 
electric charges, the fermion content of the Standard Model (SM) can be 
matched with the spin-$\frac12$ states of the $\cN\!=\!8$ supermultiplet
associated to maximally extended $\cN\!=\!8$ supergravity.
As a consequence the only extra fermions beside the usual 48 SM
spin-$\frac12$ fermions would be eight supermassive gravitinos, which
carry spin $s=\frac32$. Under the SU(3)$_c\,\times\,$U(1)$_{em}$ subgroup 
of the Standard Model gauge group these split as
\beq\label{GravCharges}
         \left({\bf 3}\,,\,\frac13\right) \oplus \left(\bar{\bf 3}\,,\,-\frac13\right)
         \oplus \left({\bf 1}\,,\,\frac23\right) \oplus \left({\bf 1}\,,\, -\frac23\right)  \; ;
\eeq
see \cite{JHEP} for more detailed arguments and for a derivation 
of these quantum numbers~\footnote{Note that there are no low
energy superpartners  in our scheme because supersymmetry 
and its associated $R$-symmetry are replaced by the conjectured 
E$_{10}$ and K(E$_{10}$) symmetries of a unified theory.}. We will here focus
on the color singlet gravitinos carrying charge $q = \pm \frac23 e$,
which are subject only to electromagnetic and gravitational 
interactions (the color triplet gravitinos play no role in our considerations).
Due to their  fractional charges the gravitinos cannot decay into 
SM fermions, and are therefore stable independently of their mass. 
Their stability against decays makes them natural DM candidates 
\cite{MeissnerNicolai2019}.  As explained there, their abundance 
cannot be determined from first principles, since they were never 
in thermal equilibrium due to their extremely small annihilation 
cross section. Nevertheless, one can plausibly assume that, 
with an equal number of positively and negatively charged gravitinos, 
their abundance is approximately given by the average DM density 
inside galaxies. From the numbers given in \cite{WdB} it then follows that, if
DM were entirely made out of nearly Planck mass particles, this would
amount to $\sim 3\cdot 10^{-13}$ particles per cubic meter within our solar system.
A more accurate estimation of flux rates is hampered by important 
uncertainties, such as possible inhomogeneities in the DM distribution 
within galaxies or stellar systems (see {\em e.g.} \cite{DK}). The velocity 
distribution is expected to be centered around the values $10^{-4} \lesssim
\beta \lesssim 10^{-2}$. Indeed, from the virial theorem we expect the 
velocity of these particles in the vicinity of the earth to be $\sim 30$ km$\,\cdot\,$s$^{-1}$ 
($\beta\sim 10^{-4}$) if they are bound to the Solar System, and $\sim 230$ km$\,\cdot\,$s$^{-1}$ 
($\beta\sim 10^{-3}$) if they are bound to our Galaxy. 

A supermassive electrically charged particle, even non-relativistic, can easily pass through 
the earth without attenuation or deflection, losing only a tiny fraction of its huge kinetic energy 
during its passage through the earth. In fact, any particle of mass greater than $10^{14}$ GeV will 
be able to penetrate the earth \cite{PDG}. Consequently, as we assume the 
Particle's mass to be close to the Planck mass, all processes
considered here are effectively independent of the Particle's mass.

Due to its electric charge the Particle would thus interact 
electromagnetically with both electrons and nuclei.
Because of its non-relativistic motion, and neglecting 
spin effects, we are effectively dealing with a point particle of quasi-infinite 
mass, and can thus perform the analysis in the gravitino's rest frame.
As we explained in \cite{MNLZ}, we therefore expect 
three kinds of detectable interactions to take place, namely

\begin{itemize}
\item Ionization
\item Electronic excitation without ionization
\item Nuclear recoil
\end{itemize}

For passage sufficiently far away from the nucleus the interaction is mainly 
between the Particle and the outer and more weakly bound electrons. 
A simple classical argument shows that, for velocities 
$\beta \lesssim 10^{-3}$ the energy imparted to individual electrons 
by the Particle is much less than the minimum ionization energy, hence no, 
or not much, ionization is expected to occur.

Closer to the center of the atom, but still sufficiently far away from the 
nucleus the Particle can induce electronic transitions in all shells
and lift electrons to higher excited states. This process  is expected to
give rise to detectable fluorescent light along the track (a `glow of photons'), 
and has been thoroughly analyzed for the JUNO detector in \cite{KLMN}, 
with specific rates also calculated for Argon (but not Xenon) based detectors.

Lastly, if the Particle moves inside the innermost electron orbit, the 
collision will look more and more like a nuclear recoil, with a clearly
visible signal if the Particle hits the nucleus head-on. 
In the Particle's rest frame the process effectively corresponds to 
scattering of a screened nucleus against 
a fixed point-like target of infinite mass and charge $q =\pm \frac23 e$,
if the Xenon atom existed in isolation. However, because the process 
takes place in a dense medium of Xenon atoms we must
resort to the Lindhard formula quoted below.

If the Particle's velocity is $\beta$ the maximal velocity that can be
imparted to the nucleus in the recoil is $2\beta$ with recoil energy
 ${\rm E}_{recoil} =2M_{\rm Xe}\beta^2$. The range of energies of 
the recoil that can be observed with LZ (with efficiency bigger than 90\%) 
is between 14$\,\keV$ and 250$\,\keV$, which for a Xenon nucleus of mass 
$\sim\,$120 GeV  gives
\beq\label{MeV}
14 \keV <  E_{recoil} < 250 \keV \; \Rightarrow \;\;
2.4\cdot 10^{-4}\,<\beta <\,  10^{-3}
\eeq
This is well within the range of our assumed velocity distribution.
In particular, the observed large value of the recoil energy agrees
very well with the assumed galaxy bound velocity of 
the DM particles.

The expected rate of events also depends on the average velocity of
the gravitinos, which, as we said, is subject to a number of uncertainties. 
With an assumed gravitino mass of the order of Planck mass, local density of
Dark Matter $0.3\cdot 10^6$ GeV$\,\cdot\,$m$^{-3}$,
and  velocity of $\lesssim$ 220 km$\,\cdot\,$s$^{-1}$, this
leads to the flux estimate \cite{MeissnerNicolai2019} 
\beq
\Phi\,\lesssim \, 0.03 \, {\rm m}^{-2} {\rm yr}^{-1} {\rm sr}^{-1}
\eeq
With a fiducial  area of LZ $\sim 1 {\rm m}^2$ we thus arrive at an
estimate for the number of gravitinos coming from all directions
\beq
\# \; \mbox{(events per year)} \; \sim \; \cO (0.3) \;\; ,  
\eeq
again modulo the mentioned uncertainties. 

It remains an open question whether there is any way to confirm 
that the Particle carries charge $\pm \frac23 e$ and spin
$s=\frac32$, via the gravitino's spin dependent coupling (\ref{Lint}).

\section{Estimates for the LZ detector}

The estimates in \cite{MNLZ} and \cite{KLMN} were mostly tuned towards 
expectations for the JUNO experiment, but also with a brief discussion of
upcoming Argon and Xenon detectors. In this section we refine these 
estimates specifically for the LZ experiment, in view of its 
next observational run and its next data release. In the final section 
we will specify what kind of extra specific signs and signals the LZ detector 
could watch out for to validate or refute our hypothesis.

To express the recoil cross section for an incoming Particle and 
an immobile Xenon atom it is better to use the recoil momentum of the atom
rather than the recoil energy since the former is 
independent of the reference frame.
The (electromagnetic) differential cross section for nuclear recoil 
as a function of recoil momentum $Q$ for a superheavy 
incoming gravitino of charge $ \pm \frac23 e$ hitting a Xenon 
nucleus of mass $M_{\rm Xe}$ and charge 54$e$ with velocity $v$ is 
given by the Lindhard formula \cite{Lindhard}
\beq  
\frac{\rd \si}{\rd Q}=\frac{\pi a^2\sqrt{Q_{max}}}{4\eps^4 Q^{3/2}}f\left(\frac{Q}{Q_{max}}\right)
\label{csdiff}
\eeq
where
\bea
Q_{max}&=&2M_{\rm Xe}v\nn\\
\eps&=&\frac{4\pi\eps_0 aM_{\rm Xe}v^2}{64e^2}\nn\\
a&=&\frac{0.885 a_0}{((2/3)^{2/3}+54^{2/3})^{1/2}}\\
f\left(\frac{Q}{Q_{max}}\right)&=&\left[1+\frac{Q_{max}^{2/9}}{\la_0^{2/3}\eps^{8/9}Q^{2/9}}\right]^{-3/2}
\eea
$a_0$ is the Bohr radius ($0.529$ \AA) and $\la_0\sim 2.6$. 
Here the function $f$ in (\ref{csdiff})  takes into account both the electron cloud screening
and the presence of surrounding atoms.
It is important to note that 
the cross section is  bigger for smaller momentum transfers so if a large momentum 
transfer happens it is expected to be accompanied by smaller 
momentum transfers in the case of multiple nuclear recoils.

Consequently, setting $f \approx1$ we get the total cross section 
for nuclear recoil with recoil momenta $Q$ 
in the interval $Q_{min} < Q <Q_{max}$
\bea
\si(Q_{min},Q_{max})\,&=&\, \int_{Q_{min}}^{Q_{max}}  
\rd Q\, \frac{\rd \si}{\rd Q}  \nn\\[2mm]
 \,&=&\,  \frac{\pi a^2}{2\eps^4}
\left[\sqrt{\frac{Q_{max}}{ Q_{min}}} - 1\right]
\label{csnr}
\eea
Plugging in the numbers for LZ ($\eps\sim 16$, $a\sim 1.2\cdot 10^{-11}$ m, $Q_{max}\sim 240$ MeV, $Q_{min}\sim 40$ MeV) we get approximately
\beq
\si(Q_{min},Q_{max})\sim 10^{-27}\ {\rm m}^2
\eeq
for collisions of gravitinos with nuclei. We leave the calculations of the 
amount  of light coming from electronic excitations and ionizations of the Xenon 
atoms for future investigation (but see \cite{KLMN} for results 
concerning liquid Argon). An independent estimate for 
the cross section could come from the geometric considerations.
Assuming the radius of the Xenon nucleus around 6 fm the geometrical cross 
section of the Xenon nucleus is of the order of $10^{-28}$ m$^2$ 
which is consistent with (\ref{csnr}) if $Q_{min}\sim 0.8\, Q_{max}$, 
corresponding to a head-on central collision of the gravitino with the Xenon atom. 
The density of Xenon  atoms in the liquid xenon is easily calculated to be around 
$1.4\cdot 10^{28}$ m$^{-3}$ so we can estimate the number of 
nuclear recoils per distance as 
\beq
P \; \sim \; 1.4\ {\rm m}^{-1}. 
\eeq
With this minimal assumption we can calculate the 
probability of $N$ recoils over length $L$, which is given by a Poisson 
distribution
\beq
p_N \;=\; \frac{(PL)^N}{N!}\re^{-PL}.
\eeq
We assume that the effective length of the LZ detector is $L\sim 1.0\,$m. Therefore our formula gives
the following values for the probability of $N$ recoils for the
passage of a {\em single}  particle through the detector:

\vspace{7mm}
\begin{tabularx}{0.4\textwidth} { 
  | >{\centering\arraybackslash}X 
  | >{\centering\arraybackslash}X 
   | >{\centering\arraybackslash}X 
     | >{\centering\arraybackslash}X 
    | >{\centering\arraybackslash}X 
  | >{\centering\arraybackslash}X | }
\hline
N=0 & N=1 & N=2 & N=3  & N=4\\
\hline
0.247 & 0.345& 0.242 & 0.113 & 0.039\\
\hline
- & 0.457& 0.321 & 0.150 & 0.052\\

\hline
\end{tabularx}

\vspace{4mm}\noindent
{\bf   Table 1: large ($\sim 250$ keV) recoil probabilities for the passage of a single 
particle through the LZ detector; the second line shows 
the probabilities under the assumption that one recoil has taken place}

Therefore the total probability that additional large recoils are present, given the
occurrence of one recoil event is around 0.54. The probability of additional, smaller than $250$ keV, recoils is close to 1. 

\vspace{4mm}

We see that once the particle hits the detector active area the probability of its 
detection through nuclear recoil is rather large. As we already explained
the main issue here is the flux of particles and not the cross section 
(in contrast to standard WIMP searches).

It is important to emphasize that for two or more sequential recoils,
the recoil energies can differ, and will usually be smaller than the maximal 
allowed value since a collision event at a small distance away from 
the center of the nucleus produces a deflection and a smaller energy transfer. 
The event reported by the LZ experiment with energy 248 keV therefore seems 
to be a head-on collision, but there may have been accompanying collisions
with smaller recoil energy that may have gone unnoticed (for  WIMP induced
events multiple recoils are practically excluded, and thus presumably 
not looked for). If multiple recoils were found with appropriate time delay 
{\em w.r.t.} the reported detection and moreover along along a straight line, 
this would be a unique and unambiguous hint that the event was due to a
gravitino, and not a WIMP.

It would be 
the distinctive experimental signature if 
a sequence of at least three nuclear recoils along a straight line was detected, 
which can point in {\em any} direction (in particular, going upwards), 
with equal calculated non-relativistic velocity between the measured
subsequent points of recoil (the energies of the subsequent recoils do not 
have to be equal since the collisions do not have to be central). It is therefore
very important to measure as precisely as possible the positions and times
of the events in the detector. There is also the possibility that
the PMTs would detect the ionizing track consistent both with the straight 
line of nuclear recoils in the central detector 
and with the reconstructed time -- that would also be a very clear 
and unmistakable signature of a  very heavy charged particle consistent 
with the DM gravitino proposed in \cite{MeissnerNicolai2019}. 

\vspace{0.5cm}
\section{Conclusions}

Obviously, the one outlier event observed by \cite{LZ1} is not enough to make 
any statement about what kind of particle was detected, nor to definitely 
rule out that it was due to some background. In this 
concluding section we spell out some further observational steps 
the LZ experiment could take in order to allow for an unambiguous
discrimination between the WIMP and our gravitino hypotheses. 
The main difference is that a WIMP will undergo only one interaction in the 
detector, and by assumption only via nuclear recoil, but not with 
electrons. By contrast, the gravitino will interact electromagnetically
with both nucleons and electrons in the detector, and thus in 
principle leave more traces of its passage through the detector.
In order to see these interactions LZ would have to watch out
also for some other signs and signals. These are:
\begin{enumerate}
\item As we showed there is a relatively high likelihood of multiple recoils along 
         the gravitino trajectory. If only recoils with maximal recoil
         momentum (energy) are recorded, secondary (and 
         tertiary)   recoils are likely to come along a straight line and 
         with smaller recoil momentum, all within a time
         span of a few microseconds. 

         Multiple recoils would also help to eliminate various 
         explanations in terms of backgrounds competing with the 
         WIMP hypothesis and to beat  the `neutrino floor' which can mask 
         WIMP events. There is then simply no neutrino
         floor because a single neutrino recoil will 
         not give rise to follow-up collisions of the type predicted here.
\item In principle, background neutrons from Uranium or Thorium fission 
         could also produce multiple recoils. The energy $E$ of a neutron to 
         produce recoil energy $T$ is given  by
         $T=4m_n M_{Xe} E(M_{Xe}+m_n)^{-2}\cos^2\theta$
         where $\theta$ is an angle of the Xenon atom track after collision. 
         For $\theta=0$ we get $E=8$ MeV  for $T=250$ keV observed 
         recoil energy \cite{LZ1}, for non-central collisions ($\theta>0$) 
         the required energy for the same $T$ is bigger.  
         The probability of neutrons of energy $E$ coming from Uranium 
         fission is given by the Maxwellian distribution 
         $\sim \sqrt{E}\exp(-E/T_m)$ with $T_m\sim 1.3$ MeV \cite{IK}. Therefore the average
         neutron energy from fission is $1.5\, T_m =1.95$ MeV, and the probability of 
         energy 8 MeV or higher is less than 1\%. Natural Xenon has seven stable 
         isotopes,  and the cross section for elastic scattering of an 8 MeV neutron 
         on all these isotopes is between 2 and 3 barns ({\em i.e.} of the order 
         of the geometric area of the xenon nucleus) with a pronounced forward peak 
         ({\em i.e.} with small momentum transfer to the Xenon nucleus) which 
         favours even bigger, hence even more improbable, energies than 8 MeV.
         Significantly, although {\em a priori} possible,  the LZ collaboration did {\em not} 
         attribute the observed event to a fast neutron background. Neutrons giving 
         rise to such large recoils are not only extremely improbable, but 
         would move with $\beta \sim 0.13$ or more, much faster than gravitinos with 
         $\beta \sim 0.001$, so gravitino induced  multiple recoils should in principle be 
         easily distinguishable from neutron induced ones.     
\item For identification it would furthermore be desirable to precisely locate 
         the recoils, and with more than two recoils check 
         whether they lie on a straight 
         line. LZ is able to measure the $(x,y,z)$ coordinates of the recoil \cite{LZ1},
         but since sequential recoils are expected to occur within a few 
         microseconds it is necessary to register them as separate
         events within the same time span (the drift time for the electrons between 
         the two scintillation signals labeled S1 and S2 in \cite{LZ1} required 
         to record an event is about 60 microseconds \cite{LZ1}).
\item The gravitino will also interact with electrons. However, with 
         its current setup, the LZ experiment discards electron recoils 
         from the outset as no such signals are expected for WIMPs.  
         A further possibility would be to look for the `photon glow'
         predicted for the JUNO experiment \cite{KLMN}, but the 
         analogous calculation for Xenon remains to be done. 
\end{enumerate}

To conclude we would like to emphasize that the confirmation of our 
hypothesis would constitute the first {\em direct observational} hint of Planck 
scale physics, pointing towards a novel Planck scale unification of matter 
with gravitation. On a very different note, we should like to mention
that in \cite{MN4} we have proposed that the emergence of
primordial black holes in the very early universe could originate from
the gravitational collapse of gravitino lumps during the early radiation
period. The confirmed existence of this new particle could thus 
lend more credence to this proposal, thus directly connecting the 
resolution of the DM problem to a very different problem of
current astrophysics.

\vspace{0.5cm}
\noindent
 {\bf Acknowledgments:} K.A.~Meissner would like to thank AEI for
 hospitality during this work.

\end{document}